\documentclass[prb,twocolumn,showpacs,floats,eqsecnum,amsmath,amssymb]{revtex4}
\usepackage[dvips]{graphicx}

\begin{document}


\title {Character of ground state of an aperiodic frustrated Josephson junction array}

\author {Y. Azizi,$^{1}$ M. R. Kolahchi,$^{1}$ and J. P. Straley$^2$}

\affiliation{$^{1}$ Institute for Advanced Studies in Basic
Sciences, Zanjan 45195-1159, Iran}

\affiliation{$^{2}$ Department of Physics and Astronomy, University of Kentucky, Lexington,
KY, 40506, U.S.A.}

\begin{abstract}
We study the energy spectrum for an aperiodic Josephson junction ladder, as a function of frustration.
Frustration is brought about by application of a transverse magnetic field, and aperiodicity is
imposed by the arrangement of plaquettes with two incommensurate areas. We study the effect of the
incommensurate plaquette areas in conjunction with that of the aperiodicity. The structure of the energy
spectrum at deep minima is shown to be described by a model that treats the plaquettes
independently. The energy spectrum is a quasiperiodic function of frustration; short range correlations in the arrangement of plaquettes have a small effect on the energy power spectrum.
\end{abstract}

\vspace{2mm} 

\maketitle
\section{Introduction}

The electronic and transport properties
 of quasiperiodic systems
are affected by their quasicrystalline structure. In one-dimensional (1D)
systems with Fibonacci order, the energy spectrum is singularly
continuous and the wave functions are neither localized nor extended. \cite{R01} Similar
exotic electronic states are predicted for the 3D icosahedral quasicrystals. \cite{R02}
The structurally
ordered icosahedral phases and their crystalline periodic approximants
show the same transport properties, implying that the local
configurational order (common in both) plays the major role. \cite{R03}
 At the same time, study
of the Fibonacci semiconductor superlattices indicates that at low enough
temperatures, the longitudinal magnetoresistance of the quasiperiodic structure
can be distinguished from that of its low order periodic approximant. \cite{R04}

In two dimensions, two types of study have been illuminating. One has been
the study of the electronic structure, the energy spectrum and wave functions,
which again has shown results similar as above, particularly in case of the
two-dimensional (2D) Penrose lattice. \cite{R05} The other kind is inspired by
the Little-Parks effect, \cite{R06} revealing the interplay of flux quantization
and free energy and manifest in the variation of critical temperature of the 2D
lattice as a function of the applied magnetic field perpendicular to its plane. \cite{R07,R08}
 Here, we consider the ground state of an aperiodic array of Josephson junctions, mainly drawing from the second type of study.

Josephson arrays give a useful context in which to study frustration.
These are composed of weakly coupled superconductors, with an interaction
between neighboring sites,
\begin{equation}
\label{joseph1}
H=-{E_J}  \sum_{<i,j>}  cos(\theta_{i}-\theta_{j}-A_{ij}),
\end{equation}
where $\theta$ is the phase of the superconducting order parameter on
the site. $A_{ij}$ describes the effect of the magnetic field; it is proportional to the
line integral of the
component of the vector potential
along the link from $i$ to $j$.  It causes the phase of the superconducting
wavefunction to vary from place to place. If we define the gauge invariant phase
differences $\Delta_{ij} = \theta_{i}-\theta_{j}-A_{ij} + 2n\pi$,
where the integer $n$ is chosen so that $-\pi < \Delta_{ij} < \pi$,
it is readily seen that the $\Delta_{ij}$ cannot all be zero, because
the directed sum of the $A_{ij}$ around a plaquette is the magnetic
flux through it.   Further, it is found that the directed sum of the
$\Delta_{ij}$ around a plaquette
sometimes differs from the magnetic flux by a multiple of $2 \pi$; we will say that these plaquettes
contain a vortex.   The ground state is specified by a configuration
of vortices.\cite{R09}  Thus these systems exhibit frustration\cite{R10},
meaning that in the ground state, the individual terms of Eq. (\ref{joseph1}) are not simultaneously at
minimum.

For a structure such as the Penrose lattice that is composed of two types of plaquettes with irrational
area ratios, the
plaquettes can never achieve their energy minima simultaneously as the magnetic field is varied, a phenomenon
known as {\it geometrical frustration}. \cite{R11}
The degree of frustration can be
characterized by a parameter $f$, which is the ratio of the
external flux in a plaquette to the quantum of flux. Only
when $f$ is an integer for all plaquettes can all junctions be simultaneously at minimum energy.

When the array is a periodic ladder and $f$ is rational, the ground state
is a superlattice of vortices. Vallat and Beck
\cite{R2} showed that for the cosine interaction the energy $E(f)$
is a uniformly continuous function of $f$.   The ground state energy $E(f)$ and
the transition temperature $T_c(f)$ are both periodic functions of
$f$, with distinct local minima where $f$ is a simple rational, showing that
commensuration is important for this case.\cite{SB}

Behrooz et {\it al.}\cite{R07} considered the problem of flux
quantization on a quasiperiodic network of thin aluminum wires.
The network consisted of parallel evenly spaced wires in one
direction, but with unevenly spaced parallel wires in the other.
The latter spacing was generated by a substitution rule, so that
it had inflation symmetry characterized by the irrational number
$\sigma=1+\sqrt{2}$. The ratio of the areas of the two kinds of plaquettes
in the quasicrystalline network was also $\sigma$. They found
that $T_c(f)$ was no longer a periodic function of the
frustration, but it did show a series of high peaks which in
magnitude were nearly the same as that of the unfrustrated
network.

We consider aperiodic ladders consisting of two types of plaquettes
with incommensurate areas, as in Fig. 1, and study the ground state energy spectrum as
a function of frustration, $E(f)$. The critical temperature gives a measure of the ground state energy, and in
this respect we, as with Behrooz et {\it al.}, find that the deep energy minima occur
when it is simultaneously true that the magnetic flux through the large and small plaquettes is approximately an integer
multiple of the flux quantum. We show that a model that treats the plaquettes independently describes
these energy minima quite well. Considering the set of energy minima in $E(f)$, we find that
a quasiperiodic function emerges, with this quasiperiodic property being dictated by the incommensurate areas
and not the aperiodicity.

The structure of the paper is as follows. In Sec. II we introduce
the Independent Plaquette Model. We begin our study of the aperiodic
ladders in Sec. III. Conclusion
comes as Sec. IV. An Appendix complements this study in that it provides a model
with cosine interaction, having an analytic {\it fractal} solution for $E(f)$.

\section{The  Independent Plaquette Model (IPM)}

Consider first a Josephson junction
ladder in a uniform perpendicular
magnetic field. The hamiltonian for this system is
\begin{eqnarray}\label{ener}
\nonumber
H=&-&{E_J}\sum_{k} [cos(\theta_{1,k}-\theta_{1,k+1})+
cos(\theta_{2,k}-\theta_{2,k+1}) +
\\
&+& cos(\theta_{1,k}-\theta_{2,k}-2\pi f x_{k})
],
\end{eqnarray}
where $\theta_{1,k}$ and $\theta_{2,k}$ are the phases of the superconducting
order parameter along the upper and lower sides of the ladder, respectively.
We have chosen the gauge of the vector potential  so that it has zero component along the horizontal junctions.
Then the effects of the applied field are represented by
$A_{ij}=\frac{2e}{\hbar c}\int_{i}^{j}A(\vec{r})\cdot\vec{dl}
= 2 \pi f x_{k}$
where $x_k$ is the $x$-coordinate of the $k$th vertical junction (a
rung of the ladder).
The frustration
factor $f$ is the  flux through a plaquette of
unit area, measured in units of the flux quantum.
\begin{figure}[ht!]
\centerline{\includegraphics[width=10cm]{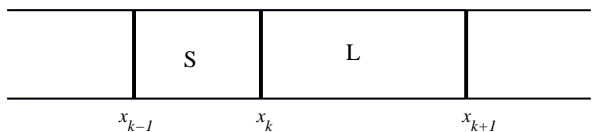}}
\vspace{0.1cm}
 \caption{Two kinds of plaquettes with different areas that are part of a Josephson junction ladder.
The arrangement of such plaquettes in the ladder can follow an
aperiodic sequence.}
\vspace{0cm} \label{ladder}
\end{figure}

For a periodic ladder, the ground state energy is a periodic function of $f$.
The energy takes on the unfrustrated value (corresponding to the peaks in $T_{c}$)
for values of $f$ that can be understood by imposing the condition that no
plaquette be frustrated.\cite{KS} We will show that the ``Independent Plaquette Model" is
also useful for the aperiodic situation.

First, let us consider a
single isolated plaquette containing four junctions;
furthermore, let us
choose the gauge
so that $A_{ij}$ has the same value for each successive junction as we
go around the plaquette.  Then we will have $A = \pi f S/2$, where
$S$ is the area of the plaquette, and the energy becomes
\begin{eqnarray}
\nonumber H =& -& cos(\theta_1 -\theta_2 -A) - cos(\theta_2 -\theta_3 - A)\\
&-&cos(\theta_3 - \theta_4 -A) - cos(\theta_4 - \theta_1 - A),
\end{eqnarray}
where $\theta_i$ are the phases at successive sites of the ring.
Analysis shows that the minima of $H$ occur for $\theta_i$ such that
$\theta_i - \theta_{i+1}  =  \pi n/2$ where $n$ is an integer; the
minimum energy per junction is given by
\begin{equation}
\label{singleene}
E = - cos(\frac {\pi}{2} (f S -
\lbrack f S \rbrack)),
\end{equation}
where $\lbrack x \rbrack$, throughout this paper, denotes the
integer closest to $x$. It is readily seen that the minimum energy
per junction
occurs when $f S$ is an integer, and that it is periodic in
$f$ with period $1/S$. Near a minimum, the cosine function is
relatively insensitive to variations in the $\theta_{i}$ away from
the optimum values.  This suggests that even in an array, when one
plaquette is close to its energy minimum it will not be greatly
perturbed by its links to other sites, nor will its presence greatly
perturb the other rings. Then we may attempt to approximate the ground state
energy per junction for an interacting system by averaging the plaquette energies
Eq. (\ref{singleene}).

In the present case, the ladder has two types of
plaquettes; $L$ with area $A_{L}$, and $S$ with area $A_{S}$. Let
$N_{Ladder}$ be the number of plaquettes in the ladder, $N_{L}$
the number of plaquettes of type $L$, and $N_{S}$ the number of
plaquettes of type $S$. Then within the IPM the energy per junction
of the ladder becomes
\begin{equation}
E_{IPM}=\frac{1}{N_{Ladder}}(N_{S}E_{S}+N_{L}E_{L}),
\end{equation}
where $E_{S}$($E_{L}$) is the energy per junction for a plaquette of
type $S$($L$). From ~\ref{singleene} we can write:
\begin{eqnarray}
E_{S}= -cos(\frac{\pi}{2}(A_{S}f-\lbrack A_{S}f\rbrack)),\\
E_{L}= -cos(\frac{\pi}{2}(A_{L}f-\lbrack A_{L}f\rbrack)).
\end{eqnarray}

It will prove convenient to choose $A_{S}$ to have unit area
(which sets the scale for $f$), and define
$\alpha={A_{L}}/{A_{S}}$ and $\nu={N_{L}}/{N_{S}}$.
Then the average spacing between vertical junctions
(the rungs of the ladder) is
\begin{equation}
a = \frac {1 + \nu \alpha}{1 + \nu},
\end{equation}
and
we can write Eqs.(2.5 -- 2.7) as
\begin{eqnarray}
\label{IPMenergy}
E_{IPM}=\frac{1}{1+\nu}(E_{S}+\nu E_{L}),\\
E_{S}= -cos(\frac{\pi}{2}(f-\lbrack f\rbrack)),\\
E_{L}= -cos(\frac{\pi}{2}(\alpha f-\lbrack \alpha f\rbrack)).
\end{eqnarray}

We will not claim that the IPM is anything more than a crude approximation.
Neighboring plaquettes are allowed to assign different currents to the rung they share;
not only is this ignored, the rung energy is counted twice in the averaging!
However, it is easy to use, and we will see that its predictions are not too
different from the numerical simulations (to which it seems to serve as a lower
bound).  We will be particularly interested in the values for $f$ for which
the energy per junction is close to the absolute minimum $-E_{J}$, and for these
cases the supercurrent on every junction is small, and the predictions of the IPM
become accurate.

\section{Study of aperiodic ladders}
Here we compare a numerical study of the spectrum of the ground
states of various frustrated Josephson junction ladders with the
predictions of the IPM. Each ladder is a sequence of two kinds of
plaquettes, $L$ and $S$. This enables us to distinguish the
importance of two
properties of the ladder: the ratio of areas of the two kinds of plaquettes,
 and the order or
particular arrangment of the plaquettes in the ladder.

The ground state energy $E(f)$ is found numerically using the Monte
Carlo (MC) simulated annealing routine.\cite{MCm}
A candidate ground state configuration is what is obtained at the
lowest temperature attempted, which we have taken to be $T=0.001$.
(By this we mean ${k_B}{T}/{E_J}=0.001$.)

 We can detect any kind of order that might exist in ground state 
energy as a function
of frustration, by
calculating the ``discrete power spectrum"\cite{dps} for $E(f)$ known for 
$N$ values of $f$
in a range $-F \le f \le F$ ($E(f)$ is an even function of $f$, so the 
range can
be doubled).  The Fourier transform of $E(f)$ is
\begin{eqnarray}
\hat{E}(\omega) &=& \frac {1}{2 F} \int_{-F}^{F} E(f) \exp (i 2\pi f \omega)
\nonumber
\\
&\approx& \frac {1}{N} \sum_{n=0}^{N-1} E(n F/N) \exp (i 2 n k/N)
\end{eqnarray}
where $\omega = k/F$ with $k=1-N/2,\ldots,N/2-1$.   Then the discrete
power spectrum is
\begin{equation}
S_{\omega}=(|\hat{E}_{\omega}|^2+|\hat{E}_{-\omega}|^2),
\end{equation}
To the extent that $E(f)$ can be represented as a sum of periodic 
components,
$S(\omega)$ will have peaks at the corresponding frequencies.

For better comparison, we define an incomplete sum over frequencies as
\begin{equation}
IS_{\omega= j/F}=\sum_{k=1}^{j}S_{k}.
\end{equation}
When at a frequency $j$ the power spectrum has a peak, the incomplete 
sum of the discrete power spectrum has a jump.
Therefore, jumps in $IS_{\omega}$ indicate peaks in $S_{\omega}$,
with the size of the jump indicating the summed weight of the peak.
In the results to be presented we used an increment of $1/13$ in 
frustration and $256$
data points, except for the case of $\sigma$Fibonacci where we had
an increment of $1/55$ and $5500$ points.

The power spectrum was
calculated in MATLAB by the FFT method. \cite{R8}

\subsection{$\tau$Fibonacci Ladder}
We start with a ladder made of two
types of plaquettes, with area ratio $\tau$ and arranged according to
the Fibonacci sequence; we call this the $\tau$Fibonacci ladder.
The plaquette areas are $A_{L}=\tau=(1+\sqrt{5})/{2}$ and $A_{S}=1$.
Denoting the structure in its $n$th step of construction by $U_{n}$,
the structure is constructed recursively by the rule
$U_{n+2}=U_{n+1}+U_{n}$, where summation means adjacent placement:
to get the structure in step $(n+2)$, put the structure in step
$(n+1)$ first, and that of step $n$, to its right. This generates
the sequence
\begin{equation}
S,L,LS,LSL,LSLLS,LSLLSLSL,...~.
\end{equation}
In the limit of an infinite structure this defines a quasicrystalline ladder
with $\nu=\tau$.

The quasiperiodic grid defining the sequence of long and short spacings for
our $\tau$Fibonacci ladder is given by
\cite{R9}
\begin{equation}\label{FGenr}
x_{n}=n+\delta+\tau^{-1} [ \tau^{-1} n + \beta ].
\end{equation}
where $\delta$ and $\beta$ are arbitrary constants, chosen to be
$\delta=0$, $\beta=-1/2$.

Within the IPM each plaquette minimizes its energy by itself; in this way the
expression for the energy becomes
\begin{equation}
\label{tauenergy}
E_{IPM}= - \frac{1}{1+\tau}cos(\frac{\pi}{2}(f-\lbrack
f\rbrack))-\frac{\tau}{1+\tau} cos(\frac{\pi}{2}(\tau f-\lbrack \tau
f\rbrack)).
\end{equation}

Eq. (\ref{tauenergy}) is composed of two terms; the first has period $\Delta f = 1$, while,
the second has period $\Delta f = 1/\tau$. Hence this function attains
its minima for those $f$ which are close to integers and which bring $\tau f$ close to an integer. These
include the Fibonacci numbers $2, 3, 5, 8, 13, 21$, but
there are other values of $f$ ($f = 6.11, 6.88,11.08, ... $
that also meet this criterion.
Even though the energy as a function of frustration is not periodic, it shows
order in its structure.  For example, the energy spectrum between the special
values of $f$ that minimize frustration is nearly symmetric about the midpoint;
e.g. $E(5+x) \simeq E(8-x)$.

The numerical calculation of the ground state energy is given in Fig. 2 along with
the predictions of Eq. (\ref{tauenergy}). It is not surprising that $E_{IPM} < E_{MC}$, because the IPM neglects
the interaction between the plaquettes. However, most features of the numerical curve are reproduced by the IPM. We can quantify this comparison by
calculating the incomplete integral of power spectra, shown in Fig. 3.

\begin{figure}[ht!]
\centerline{\includegraphics[width=10cm]{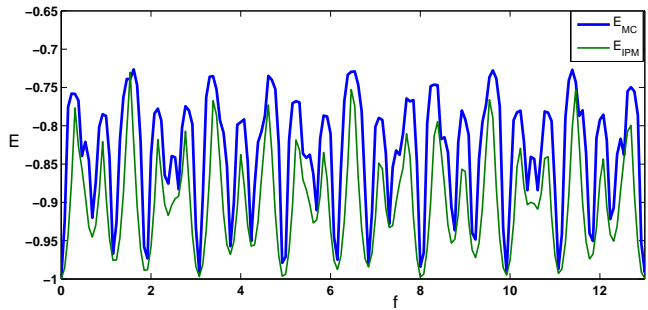}}

\vspace{0.1cm}
 \caption{Energy for the $\tau$Fibonacci ladder.  The area ratio is $\tau = (1+\sqrt{5})/2$ and the
order of the plaquettes is determined by the Fibonacci sequence.
The upper curve is the ground state energy as determined by simulated
annealing; the lower curve is the result of the Independent Plaquette Model (IPM)
approximation.}
\vspace{0cm} \label{fig1FF}
\end{figure}
\begin{figure}[ht!]
\centerline{\includegraphics[width=9cm]{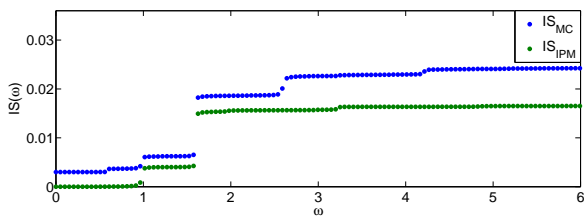}}

\vspace{0.1cm}
 \caption{Incomplete sum of the Fourier spectrum of $E(f)$ for the $\tau$Fibonacci ladder using the numerical minimization (top) and the IPM (bottom).
The top figure shows jumps at $\omega=\tau-1,1,\tau,\tau+1$.This shows that the energy is quasiperiodic as a function of $f$,
with periodicities $\Delta f = \tau^{-2},\tau^{-1}, 1,$ and $\tau$,
and weaker periodicities at various integer combinations of unity
and $\tau$. Figures are shifted relative to each other, for clarity.}
\vspace{0cm} \label{fig2FF}
\end{figure}

Both $E(f)$ contain significant periodicity corresponding to the
frequencies $\omega=1$ and $\omega=\tau$.
While the IPM energy expression is a sum of two periodic functions
with incommensurate periods, the numerical result indicates that
this function is quasiperiodic too, as it also includes the sum and
difference of the frequencies in the power spectrum: $\tau +1$ and
$\tau-1$. This is also what is found for the tight binding hamiltonian model on a
Penrose lattice.\cite{eigen}

\subsection{$\sigma$Fibonacci ladder}
We have seen that the quasiperiodic nature of the sequence and the
special ratio of the areas both play roles in the power
spectrum.  To distinguish the role of the area ratio
from the effects of the sequence we
will keep the Fibonacci order for the arrangement of the $S$ and $L$
plaquettes, but choose the areas to be in the silver ratio:
$A_{L}=\sigma=1+\sqrt{2}$ and $A_{S}=1$. The general formula
 for generating this ladder is now \cite{R9}
\begin{equation}
x_{n}=n+\delta+{[ \frac{n}{\tau}+\beta ]}\sqrt{2},
\end{equation}
with $\delta$ and $\beta$ being arbitrary, chosen here to be
$\delta=0$, $\beta=-1/2$.

For this ladder the energy within the IPM is expressed as,
\begin{equation}
E_{IPM}=-\frac{1}{1+\tau}cos(\frac{\pi}{2}(f-\lbrack
f\rbrack))-\frac{\tau}{1+\tau} cos(\frac{\pi}{2}(\sigma f-\lbrack
\sigma f\rbrack)).
\end{equation}

Figure 4 shows the result of a simulated annealing Monte Carlo study of the
ground state energy for a ladder with 144 plaquettes, along with the
predictions of the IPM, Eq. (3.5). The properties mentioned for the
$\tau$Fibonacci ladder hold here too: there are minima near $f = 2, 3, 5, 7, 10,$ and $12$ because for these values $\sigma f$ is also close to an integer.
\begin{figure}[ht!]
\centerline{\includegraphics[width=10cm]{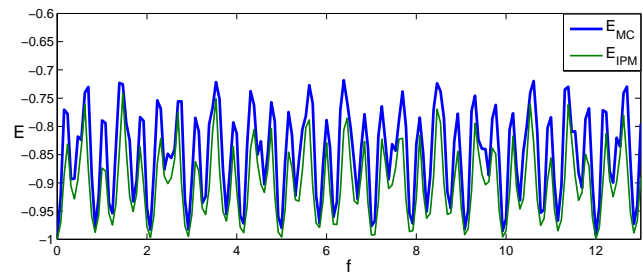}}
\vspace{0.1cm}
 \caption{Energy for the $\sigma$Fibonacci ladder.  The area ratio is $\sigma = 1+\sqrt{2}$ and the
order of the plaquettes is determined by the Fibonacci sequence. The
upper curve is the ground state energy as determined by simulated
annealing; the lower curve is the result of the Independent
Plaquette Model approximation.} \vspace{0cm} \label{fig1EF}
\end{figure}
The comparison of the Fourier spectra for the two energy plots, Fig. 5,
indicates that the dependence of the energy on $f$ is dictated by $\sigma$ and not by the
specific way the plaquettes are arranged. The ratio of the areas causes the frustration, and the energy minima occur for the values of $f$ that give nearly integer flux through each plaquette.
\begin{figure}[ht!]
\centerline{\includegraphics[width=9cm]{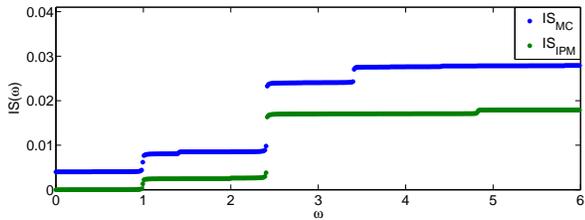}}
\vspace{0.1cm}
 \caption{Similar to Fig. 3 but when $E(f)$ is for the $\sigma$Fibonacci ladder.
The jumps at $\omega=\sigma-1,1,\sigma,\sigma+1$ show that the energy is quasiperiodic as a function of $f$,
with periodicities $\Delta f = (\sigma+1)^{-1},\sigma^{-1}, 1,$ and
$(\sigma-1)^{-1}$, and weaker periodicities at various integer
combinations of unity and $\sigma$. Figures are shifted relative to each other
for clarity. } \vspace{0cm} \label{fig2EF}
\end{figure}
The quasiperiodic nature of $E(f)$ indicates that there is a fine structure
to it that corresponds to a `many tile effect,' pointing to the presence of
vortex lattices commensurate with the underlying structure.\cite{R6, R04} The IPM cannot predict this effect,
and in Fig. (5) we
have peaks that are not reproduced by the independent plaquettes approximation, just as in the case of Fig. (3).

\subsection{Other $\tau$Ladders}

To further study the importance of plaquette order, we have  considered three ladders with $\nu=1$ and $\alpha=\tau$,
having different arrangements of the $S$ and $L$ plaquettes:
random order, the Thue--Morse sequence,\cite{tm} and a periodic lattice.

\begin{figure}[ht!]
\centerline{\includegraphics[width=9cm]{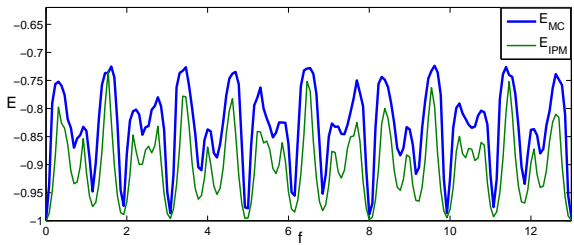}}
\vspace{0.1cm}
 \caption{Energy for the $\tau$Random ladder.  The area ratio is $\tau = (1+\sqrt{5})/2$ and the
order of the plaquettes is determined by the random sequence. The
upper curve is the ground state energy as determined by simulated
annealing; the lower curve is the result of the Independent
Plaquette Model (IPM) approximation.} \vspace{0cm} \label{fig1RP}
\end{figure}

\begin{figure}[ht!]
\centerline{\includegraphics[width=9cm]{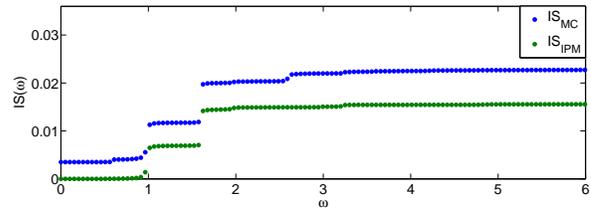}}
\vspace{0.1cm}
 \caption{Periodicity of E(f) for the $\tau$Random
ladder, as represented by the
incomplete integral of the power spectrum (similar
to Fig. 3).
It shows jumps at $\omega=\tau-1,1,\tau,\tau+1$.
This shows that the energy is quasiperiodic as a function of $f$
with periodicities $\Delta f = \tau^{-2}, \tau^{-1}, 1,$ and $\tau$,
and weaker periodicities at various integer combinations of unity
and $\tau$. The characteristics in this figure are nearly identical
with those of the other two $\tau$Ladders. Figures are shifted
relative to each other, for clarity.} \vspace{0cm} \label{fig2RP}
\end{figure}

For the random ladder, the results for a simulation using 144
plaquettes is shown in
Fig. \ref{fig1RP}.  According to our simulated annealing study, there is less fine structure in the random ladder than
predicted by the IPM.  It remains true that the IPM consistently
gives a slightly lower energy that the simulated annealing,
and that it locates the energy minima. The incomplete integral of the power spectrum of the
$\tau$Random $E(f)$ is shown
in Fig. \ref{fig2RP}; the other two $\tau$Ladders mentioned here
give similar results.

We have also studied the $\tau$Periodic lattice (a periodic sequence of $SL$, with areas having
ratio $\tau$). as well as one
having the Thue-Morse order (built recursively by appending
the complement of a sequence to itself: $SL, SLLS, SLLSLSSL,...$.  ).  Our results are summarized in Table \ref{table1} for the
numerical results and Table \ref{table2} for the IPM results. We see that all lattices have peaks at $\omega = 1$, $\alpha$ and $\alpha \pm 1$, and this shows that $E(f)$ is a quasiperiodic function with main peaks at $\omega = 1$ and $\alpha$. From these results we separate out the role of the areas of plaquettes from that of the order in lattice. $\tau$Fibonacci and $\sigma$Fibonacci lattices have the same order but with different areas of plaquettes. The results for these two lattices show that the position of the main peaks is determined by the areas of plaquettes.


We can further justify the results by a quantitative argument that starts with the IPM model.
From the IPM model we know that the main peaks in $S(\omega)$ are proportional to the square
of the density of $S$ and $L$. These are just the coefficients of $E_S$ and $E_L$, in the expression
for energy (Eqs. 2.8--2.10).
Then the peak heights depend only on $\nu$; this is why the last three rows of Table II
(for different $\nu = 1$ models) have identical
entries for the corresponding peaks.
In the same way, we can explain the added peaks
in the MC data of Table I  in terms of the correlations between plaquette types.
We  define $D_{LS}$ to be the probability that the two types of plaquettes
appearing adjacent to each other: the ratio of all possible $LS$ and $SL$ pairs to the total possible number of pairs in the
  array.   This is correlated with the height of the peaks at $\omega = \alpha+1$, as shown in Table III.
For the $\tau$ ladders, the peak height is proportional to $D_{LS}$ to good approximation. There is a similar explanation\cite{aziziunp} for
the amplitudes of the peaks at $\omega = \tau-1$ that uses the $LS$ combination
as its basic unit, so that two--plaquette interactions account for this case, too.



\begin{table}
\begin{tabular}{|c|c|c|c|c|}
  \hline
  $\tau$Fibonacci & $\tau$-1 (0.653) & 1 (2.360) & $\tau$ (12.14) & $\tau$+1 (3.518) \\ \hline
  $\sigma$Fibonacci & 1 (3.457) & $\sigma$-1 (0.397) & $\sigma$ (14.0) & $\sigma$+1 (3.249) \\ \hline
  $\tau$periodic & $\tau$-1 (0.565) & 1 (7.132) & $\tau$ (8.111) & $\tau$+1 (3.899) \\ \hline
  $\tau$Thue-Morse & $\tau$-1 (0.422) & 1 (10.16) & $\tau$ (9.387) & $\tau$+1 (2.334) \\ \hline
  $\tau$Random & $\tau$-1 (0.5285) & 1 (7.138) & $\tau$ (8.164) & $\tau$+1 (1.455)  \\
  \hline
\end{tabular}
  \centering
  \caption{Peak locations and heights for different lattices, according to the simulation. For each lattice four main peaks given; number in parentheses gives height of peak (we multiply it by 1000 for better representation). }\label{table1}
\end{table}
\begin{table}
\begin{tabular}{|c|c|c|}
  \hline
  $\tau$Fibonacci & 1 (4.208) & $\tau$ (11.008)   \\ \hline
  $\sigma$Fibonacci & 1 (4.208) &  $\sigma$ (11.008)  \\ \hline
  $\tau$periodic  & 1 (7.2) & $\tau$ (7.2)  \\ \hline
  $\tau$Thue-Morse & 1 (7.2) & $\tau$ (7.2)  \\ \hline
  $\tau$Random  & 1 (7.2) & $\tau$ (7.2)   \\
  \hline
\end{tabular}
  \centering
  \caption{Peak locations $\omega$ and heights (in parentheses) for different lattices, according to the IPM. For each lattice two main peaks given; number in parentheses gives height of peak (we multiply it by 1000 for better representation). }\label{table2}
\end{table}
\begin{table}
\begin{tabular}{|c|c|c|}
  \hline
  $\tau$Fibonacci & 2/$\tau^2$ & 6.028  \\ \hline
  $\sigma$Fibonacci & 2/$\tau^2$ &  5.567  \\ \hline
  $\tau$periodic  & 1 & 5.899 \\ \hline
  $\tau$Thue-Morse & 2/3 & 5.252 \\ \hline
  $\tau$Random  & 1/2 & 5.820   \\
  \hline
\end{tabular}
  \centering
  \caption{The ratio of all possible $LS$ and $SL$ pairs to the total possible number of pairs in the
  array is defined as $D_{LS}$, and given in first column. The second column gives the ratio of $S(\alpha+1)$ to $D_{LS}^2$ (we multiply it by 1000 for better representation).}\label{table3}
\end{table}

In summary, our study of the one dimensional lattices shows that
$E(f)$ is a quasiperiodic function with the main frequencies at
the areas of the plaquettes. This property is an extension of
that of the periodic behavior for the square lattice.\cite{lobb}
We can also observe an approximate mirror symmetry in the dependence
of $E$ on $f$; by that we
mean, there is a set of values $f_n$ for which the energy is
very close to its unfrustrated value, and between these values
$E(f)$ has a mirror symmetry about $f=(f_n+f_{n+1})/2$.
  This can again be viewed as an extension of the mirror symmetry for the
square lattice.

\section{conclusion}
The independent plaquette model treats the plaquettes as if they
were independent of each other, and at the deep energy minima, this
nearly coincides with the energy minimum when such an approximation
is lifted.

By using the numerical minimization of Hamiltonian \ref{ener} we
find some interesting results about the structures of ground state
energy $E(f)$. We see that $E(f)$ has two main frequencies and their
ratio equals the ratio of areas of the two plaquettes. There are other frequencies in the spectrum of $E(f)$ equal to sum and
difference of the two main frequencies, indicating that $E(f)$ is a
quasiperiodic function and not just a sum of two periodic functions
with incommensurate periods.
The order of the lattice affects the amplitudes
of different frequencies in the spectrum of $E(f)$. Study of the energy
power spectrum gives a role to the lattice ordering because this affects
the correlations between $S$ and $L$, which goes
beyond the IPM. However, this is only a short range effect; we have no
evidence that the long range
order in the array plays any role.

If we look at the behavior of $T_c(f)$, which roughly goes hand
in hand with the ground state energy, the linearized mean-field
theory provides an analytical approach to the
problem.\cite{stroud} The results of this model show that
$T_c(f)$ is a quasiperiodic function.\cite{qnori} In their
respective Fourier spectra, we observe peaks on the same
 special combinations of the main
frequencies, and this is an indication of the particular order
present in the lattice.

The independent plaquette model provides an excellent approximation
for $E(f)$, following nearly all its features, while
apparently giving a lower
bound for it.

\section*{Acknowledgement}
Y.A. and M.R.K. acknowledge the support of a grant from the Institute for Advanced Studies
in Basic Sciences.

\renewcommand {\theequation} {A-\arabic{equation}}
\setcounter{equation}{0}
\section*{APPENDIX: IPM and the  $J^2$ model}

In a study with similar motivation, Grest, Chaikin and
Levine \cite{R4} (GCL) numerically studied the ground states of a one
dimensional quasiperiodic array with inflation symmetry.
Although their model was inspired by the Josephson network, it differed from it
in two ways:

*They simplified the form of the interaction, in effect replacing $cos(x)$ by
$1 - \frac{1}{2} (x-2 \pi n)^2$ (where $n$ is the integer that gives the minimum
value for this expression\cite{Jnote}).  In practice the arguments of the cosine function
are small (mod $2 \pi$), so that this is a good approximation; it somewhat changes
the energy cost of introducing a vortex, but this is unimportant since the number of
vortices is set by the applied field.
They refer to this as the ``current squared" ($J^{2}$) model.

*They assumed that no current is carried along the periodic wires. \cite{noteJ2}  This forces
the phases along these wires to be all the same (in a certain gauge), and implies
that all plaquettes along a column either all contain a vortex or all do not.
This
reduces the mathematics problem to one dimension, so that
 Griffiths and
Flor\'{\i}a \cite{R5} (GF) were able to solve the model exactly.
In their numerical study, GCL found that there are deep minima in
$E(f)$ which they interpreted as due to a kind of commensuration between
the current pattern and the underlying lattice.
Similar results were found in experimental studies by  Chaikin et al. \cite{R6}
For the $J^2$ model,
GF showed that $E(f)$ takes on the value
$1-{\pi^{2}}/{3}$
(the averaged value of the approximated cosine function)
for almost all $f$, but that
there is a countable set of irrational values of $f$
at which $E(f)$ takes on a lower value.

We have some misgivings about the choice of model.
The way that GCL reduce it to one dimension does some violence to
the physical situation.
 A column of vortices is unstable, because the vortices
repel each other.  We would expect that the ground state configurations would
have the vortices more uniformly spaced in two dimensions.
Having no current on the periodic wires also means
that the current is the same on all of the nonperiodic wires and does not vary
along it.  We will show that with this very strong constraint, it is not
necessary to make the $J^2$ approximation: the model can be exactly solved
for the cosine interaction (with very little difference in the results).

The GCL model is defined by the energy function
\begin{equation}
H = -E_{J} \sum_{i,j} [cos(\theta_{i,j}-\theta_{i+1,j})
+ cos(\theta_{i,j} - \theta_{i,j+1} - 2\pi f x_{i}) ] .
\end{equation}
As an approximation, GCL assumed  $\theta_{i,j}$ to be independent of $i$, which
implies that there is no current along the ``horizontal" links corresponding
to the index $i$.  This makes the first term a constant which will not
be included in what follows.
Minimizing the energy with respect to $\theta_{j}$
gives
\begin{equation}
sin(\theta_{j}-\theta_{j+1} - 2\pi f x_{i}) = sin(\theta_{j-1}-\theta_{j}
-2\pi f x_{i}),
\end{equation}
which implies $\theta_{j}-\theta_{j+1} =\theta_{j-1}-\theta_{j}$  (mod($2 \pi$)).
Then there is just one undetermined variable: the difference in phase
$b$
between adjacent columns.  Specification of $b$ determines all of the currents
and link energies in the form
\begin{equation}
\label{GFenergy}
E = - \lim_{N \rightarrow \infty} \frac {1}{N} E_{J} \sum_{j=1}^{N}
 cos(b - 2 \pi f x_{j}).
\end{equation}

This is not the same as the model that we have been discussing, but the
results can be compared to those of the IPM.
Applying the IPM to the two-dimensional array again leads to Eqs. (2.8 -- 2.10)
since it decouples neighboring plaquettes.

Eq. (\ref{GFenergy}) can be evaluated exactly following the methods of GF.
They define
$r(y)$ to be the distribution of the values $y_{j} =  f x_{j} $ mod(1),
so that the energy can be calculated
from
\begin{equation}
E =- E_{J} \int_{0}^{1}   cos(2 \pi y) r(y-b) dy.
\end{equation}
They consider sequences $\{x_{j}\}$ that can be written in the form
$x_{i} = a [i + g(i \omega)]$ where $g(x)$ is the periodic extension (with period 1)
of the function $g(x) = \gamma x$.  These include the
 periodic, $\tau$, and $\sigma$ ladders ---
for example, the $\tau$ ladder corresponds to
$a = 3-\tau$, $\gamma = 1/\sqrt(5)$, and $\omega = \tau -1$.
They show that for almost all values of $f$, $r(y)$ is the uniform
distribution $r(y) = 1$ (so that $E = 0$),
but that $r(y)$ has nontrivial form
for
$af = (l + m \omega)/n$, where $l$,
$m$, and $n$ are integers and $a$ sets the scale.  The effects for
$n > 1$ are relatively
small and will not be considered further.  Then

\begin{eqnarray}
r(y) &=& (K+1)/\kappa ~\quad\hbox{for}\quad -s/2 < y < s/2,\\
 &=& K/\kappa ~~~~~~~~~\quad\hbox{otherwise}.
\end{eqnarray}
Here, we have $s=\kappa - K$, with
$\kappa = | (1 + \gamma \omega)m + \gamma l |$,
and $K$ the largest integer less than $\kappa$.
The corresponding energy is\cite{levstein}
\begin{equation}
\label{cosenergy}
E = - E_{J}  |sin (\pi \kappa)/\pi \kappa |.
\end{equation}
Because GF follow
GCL making the approximation $cos(x) = 1 - x^{2}/2$, they obtain a slightly
different result.

It is readily seen that the
most negative values for Eq.(\ref{cosenergy}) occur when $\kappa$ is small,
i.e. when $m/l \approx |1 + \gamma \omega|/|\gamma|$.
The dependence of the energy  on $f$ described by Eq.(\ref{cosenergy}) is a discontinuous fractal function
--- a forest of vertical lines, rather than the smooth curves we find numerically
for the Josephson ladder.  We believe this to be a real difference between the two
models.
In any case, the two results agree on the positions of
the deepest minima.


\begin{references}
\bibitem{Jnote} References 19, 25, and 28 chose to define their models in
terms of functions of unit period, whereas we use functions with period
$2 \pi$.  We have translated their results into our notation.
\bibitem{noteJ2} This approximation is not a necessary part of the $J^{2}$
model.
\end{references}
\end{document}